# Unlimited Energy Gain in the Laser-Driven Radiation Pressure Dominant Acceleration of Ions

S. V. Bulanov[1,2], E. Yu. Echkina[3], T. Zh. Esirkepov[1], I. N. Inovenkov[3], M. Kando[1],

F. Pegoraro[4], and G. Korn[5]

[1]Kansai Photon Science Institute, JAEA, Kizugawa, Kyoto 619-0215, Japan
[2]A. M. Prokhorov Institute of General Physics of Russian Academy of Sciences,
Moscow 119991, Russia
[3]Faculty of Computational Mathematics and Cybernetics, Moscow State University,
Moscow 119899, Russia
[4]Physical Department, University of Pisa and CNISM, Pisa 56127, Italy
[5]Max Plank Institute of Quantum Optics, Garching 85748, Germany

## Abstract

The energy of the ions accelerated by an intense electromagnetic wave in the radiation pressure dominated regime can be greatly enhanced due to a transverse expansion of a thin target. The expansion decreases the number of accelerated ions in the irradiated region increasing the energy and the longitudinal velocity of remaining ions. In the relativistic limit, the ions become phase-locked with respect to the electromagnetic wave resulting in the unlimited ion energy gain. This effect and the use of optimal laser pulse shape provide a new approach for great enhancing the energy of laser accelerated ions.

# 1. Introduction

The radiation pressure of a super-intense electromagnetic pulse on a thin quasi-neutral plasma slab has been proposed in Ref. [1] as an acceleration mechanism able to provide ultrarelativistic ion beams. In this radiation pressure dominant acceleration (RPDA) regime (also called the "Laser Piston" or the "Light Sail" regime), the ions move forward with almost the same velocity as the electrons and thus have a kinetic energy well above that of the electrons. This acceleration process is highly efficient, with the ion energy per nucleon being proportional in the ultrarelativistic limit to the electromagnetic pulse energy. The idea of transferring momentum from light to macroscopic objects goes back to [2]. In the mid '50s of the last century ion acceleration by a high intensity electromagnetic wave incident on an electron cloud carrying a small portion of ions was considered by V.I. Veksler [3] for conditions when the ion acceleration occurs in the collective electric field which is produced due to the radiation pressure acting on the electron component. An analytical description of a charged particle dynamics under the radiation pressure can be found in Ref. [4] (chapter 9, problem 6), where a solution is obtained for the motion of a charge under the action of the average force exerted upon it by the wave scattered by it. There is an analogy between the RPDA mechanism and the "Light Sail" scheme for spacecraft propulsion. This scheme, which uses the photon momentum transfer to the light-sail, has been proposed by F. A. Zander in 1924 [5]. The use of lasers for propelling the light-sail over interstellar distances has been proposed in Ref. [6]. (for details and further discussions see Ref. [7]).

Recently the RPDA regime of laser ion acceleration has attracted great attention (e.g. see review article [8]). In Refs. [9, 10] the stability of the accelerated foil has been analyzed. Refs. [11, 12] are devoted to extending its range of operation towards lower electromagnetic wave intensities. The interaction of a high intensity laser pulse with extended plasmas in the RPDA (or 'Laser Piston') regime has been simulated in [13]. In Refs. [1, 14] effects of the foil transparency are considered. A foil accelerated to relativistic energies by a laser pulse can act as a relativistic flying mirror for frequency up-shift and intensification of a reflected counter-propagating light beam [15]. An indication of the effect of the radiation pressure on bulk target ions is obtained in experimental studies of plasma jets ejected from the rear side of thin solid targets irradiated by ultraintense laser pulses [16].

While publications develop regimes of energy enhancement of the accelerated ions by exploiting the dependence on the pulse polarization of the laser interaction with matter [11] and target structuring [17], in the present paper we propose to use targets expanding transversally in order to increase the energy of accelerated ions. The transverse expansion of the accelerated shell can be provided by the action of the ponderomotive force of a laser pulse with a finite waist. It can also occur as a result of the instability described in Ref. [9].

## 2. Mathematical model

The nonlinear dynamics of a laser accelerated foil is described within the framework of the thin shell approximation first formulated by E. Ott [18] and further generalized in Refs. [9, 19]. In the electromagnetic wave interaction with a thin foil, the latter is modeled as an ideally reflecting mirror. The equations of motion of the surface element of an ideally reflecting mirror in the laboratory frame of reference can be written in the form

$$\frac{d\boldsymbol{p}}{dt} = \frac{\mathcal{P}\boldsymbol{v}}{\sigma}, \qquad (1)$$

where $\boldsymbol{p}, \mathcal{P}, \boldsymbol{v}$, and $\sigma$ are the momentum, light pressure, unit vector normal to the shell surface element, and surface density, $\sigma = nl$, respectively. Here $n$ and $l$ are the plasma density and shell thickness. The geometry is illustrated in Fig. 1. We determine the surface element $\Delta s$ as being carrying $\Delta N = \sigma \Delta s$ particles, which is constant in time. We take the shell initially to be at rest, at $t = 0$, in the plane $x = 0$. In order to describe how its shape and position change with time it is convenient to introduce the Lagrange coordinates $\eta$ and $\zeta$ playing the role of the markers of the shell surface element. The shell shape and position are given by the equation

$$\boldsymbol{M} = \boldsymbol{M}(\eta, \zeta, t) \equiv \{x(\eta, \zeta, t), y(\eta, \zeta, t), z(\eta, \zeta, t)\}. \qquad (2)$$

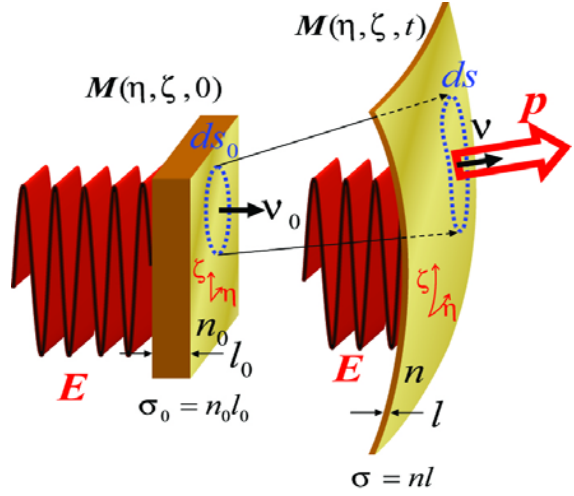

Figure 1. Evolution of thin shell irradiated by strong electromagnetic wave.

At a regular point, the surface area of a shell element and the unit vector normal to the shell are equal to

$$\boldsymbol{\nu}\Delta s = \partial_\eta \boldsymbol{M} \times \partial_\zeta \boldsymbol{M}\, \mathrm{d}\eta \mathrm{d}\zeta \qquad (3)$$

and

$$\boldsymbol{\nu} = \frac{\partial_\eta \boldsymbol{M} \times \partial_\zeta \boldsymbol{M}}{|\partial_\eta \boldsymbol{M} \times \partial_\zeta \boldsymbol{M}|}, \qquad (4)$$

respectively (e. g. see [20]). The particle number conservation implies $\sigma \Delta s = \sigma_0 \Delta s_0$, where $\sigma_0 = n_0 l_0$. This yields

$$\sigma = \frac{\sigma_0}{|\partial_\eta \boldsymbol{M} \times \partial_\zeta \boldsymbol{M}|}. \qquad (5)$$

Using these relationships we obtain the equations of motion

$$\sigma_0 \partial_t p_i = \mathcal{P} \varepsilon_{ijk} \partial_\eta x_j \partial_\zeta x_k, \tag{6}$$

$$\partial_t x_i = c \frac{p_i}{(m_\alpha^2 c^2 + p_k p_k)^{1/2}}. \tag{7}$$

Here $m_\alpha$ is the ion mass, $\varepsilon_{ijk}$ is the unit antisymmetric tensor, $i = 1, 2, 3$, and summation over repeated indices is assumed.

The radiation pressure on the shell exerted by a circularly polarized electromagnetic wave propagating along the $x$-axis with amplitude $E = E(t - x/c)$ is

$$\mathcal{P} = \frac{E^2}{2\pi} \left( \frac{1 - \beta}{1 + \beta} \right) \tag{8}$$

Here $\beta = p_x (m_\alpha^2 c^2 + p_x^2)^{-1/2}$ is the shell normalized velocity.

### 3. Ion acceleration in the expanding foil

We consider the case when the accelerated shell moves in the longitudinal direction with an ultrarelativistic velocity, i. e. $p_x / m_\alpha c \gg 1$. The shell expands or contracts in the transverse directions with momentum components satisfying the conditions $p_y / p_x \ll 1$ and $p_z / p_x \ll 1$, i. e. the transverse momentum is relatively small compared to the longitudinal momentum. Using this condition, we look for solutions of Eqs. (6, 7) assuming a local dependence of the transverse components of the coordinates and of the momentum in the form

$$y = \Lambda_y(t)\eta, \qquad z = \Lambda_z(t)\zeta, \tag{9}$$

which corresponds to local maxima or minima in the $x(t)$, where $x(t)$ is locally independent of $\eta$ and $\zeta$. In this case the right hand side of Eq. (6) for the transverse components of momentum vanishes. The shell expands (contracts) ballistically with

$$p_y = \pi_y^0 \eta, \qquad p_z = \pi_z^0 \zeta,$$

$$\Lambda_y(t) = 1 + \frac{\pi_y^0}{m_\alpha} \int_0^t \frac{dt'}{\gamma(t')}, \quad \Lambda_z(t) = 1 + \frac{\pi_z^0}{m_\alpha} \int_0^t \frac{dt'}{\gamma(t')}. \tag{10}$$

Here $\pi_y^0$ and $\pi_z^0$ are constant and $\gamma(t)$ is the relativistic gamma-factor of the shell. Within the framework of our approximation it is given by $\gamma = (1 + (p_x / m_\alpha c)^2)^{1/2}$. The areal density changes as

$$nl = \frac{n_0 l_0}{\Lambda_y \Lambda_z}. \tag{11}$$

Inserting these expressions into the longitudinal component of Eq. (6), we obtain the equation for $p_x$:

$$\frac{dp_x}{dt} = \frac{m_\alpha v_E^2}{l_0} \left( \frac{1-\beta}{1+\beta} \right) \Lambda_y \Lambda_z, \tag{12}$$

where

$$v_E^2 = \frac{E^2}{2\pi n_0 m_\alpha}. \tag{13}$$

In the case of no expansion of the shell we have $\pi_y^0 = 0$, $\pi_z^0 = 0$, and $\Lambda_y = \Lambda_z = 1$, and Eq. (12) gives (for a constant amplitude pulse) the following asymptotic time dependence of the particle momentum [1] for $t \to \infty$

$$p_x(t) = m_\alpha c \left( \frac{t}{\tau_{1/3}} \right)^{1/3}, \tag{14}$$

i.e. $p_x \propto t^{1/3}$, which is similar to the dependence on time of the momentum of a charged particle accelerated by the light radiation pressure [4]. Here $\tau_{1/3} = 4l_0 c / 3v_E^2$.

In the case of transverse expansion along the y-axis only ($\pi_y^0 > 0$, $\pi_z^0 = 0$) at $t \to \infty$

$$p_x(t) = m_\alpha c \left(\frac{t}{\tau_{1/2}}\right)^{1/2}, \qquad (15)$$

with $\tau_{1/2} = \left(m_\alpha c l_0 / v_E^2 \pi_y^0\right)^{1/2}$. According to (11) the shell areal density decreases as $nl \propto t^{-1/2}$.

As found in Ref. [10], in the case of transverse expansion ($\pi_y^0, \pi_z^0 > 0$), in the limit $t \to \infty$, Eq. (9) yields

$$p_x(t) = m_\alpha c \left(\frac{t}{\tau_{3/5}}\right)^{3/5}, \qquad (16)$$

where $\tau_{3/5} = \left(48 l_0 m_\alpha^2 c / 125 v_E^2 \pi_y^0 \pi_z^0\right)^{1/3}$. The shell areal density decreases as $nl \propto t^{-4/5}$. We see that the momentum, $p_x \propto t^{3/5}$, grows faster than in the previous cases of a non-expanding shell or a shell expanding along one transverse direction.

Now we consider a laser pulse of finite length and assume its amplitude to be constant $E = E_0$ in the interval $0 < t - x/c < t_{las}$ and be equal to zero for $t - x/c < 0$ and for $t - x/c > t_{las}$. Introducing the wave phase as a new independent variable

$$\psi = \omega_0 \left(t - \frac{x}{c}\right) = \omega_0 \int_0^t (1 - \beta(t')) dt'. \qquad (17)$$

we find that in the limit $p_x \to \infty$ Eqs. (16, 17) yield

$$\psi \approx \text{const} + \frac{m_\alpha^2 c^2 \omega_0}{2} \int^t \frac{dt'}{p_x^2(t')}. \tag{18}$$

We can see that in the case of a non-expanding shell, when the momentum dependence on time is given by Eq. (14), the integral in the right hand side of expression (18) diverges, while for an expanding shell it follows from expressions (16) and (17) that the phase shift between the laser pulse and the accelerated ions remains finite. For a long enough laser pulse the ions at the pulse axis become trapped inside the pulse with their energy formally growing unlimitedly at the expense of the particle number decrease. We notice here that the unlimited electron acceleration regimes are well known for the electrons accelerated by the electromagnetic wave in the cyclotron autoresonance regime [21], when electrons are trapped by electrostatic wave propagating perpendicularly to the magnetic field [22] and in inhomogeneous plasmas with a downgrading density [23].

Substituting the dependence of the ion momentum on time in the form $p_x(t) = m_\alpha c (t/\tau_k)^k$ into the integral in r.h.s. of Eq. (17), we obtain

$$\psi = \omega_0 t - \omega_0 \tau_k \frac{(t/\tau_k)^{1+k}}{1+k} {}_2F_1\left(\frac{1+k}{2k}, \frac{1}{2}, \frac{1+3k}{2k}; -\left(\frac{t}{\tau_k}\right)^{2k}\right), \tag{19}$$

where ${}_2F_1(\alpha, \beta, \gamma; z)$ is the Gauss hypergeometric function [20]. Asymptotically expression (19) yields for $t \to \infty$

$$\psi \to \omega_0 \tau_k \frac{(t/\tau_k)^{1-2k}}{2-4k} + \frac{\omega_0 \tau_k}{\pi^{1/2}} \Gamma\left(\frac{2k-1}{2k}\right) \Gamma\left(\frac{k+1}{2k}\right), \tag{20}$$

where $\Gamma(z)$ is the gamma function [20]. If the power index $k$ is larger than $1/2$, the first term in the r.h.s. of Eq. (16) tends to zero for $t \to \infty$ and the phase $\psi$ remains finite being equal to the second term. When $k = 1/2$, Eq. (17) yields

$$\psi = \omega_0 \left[ t - \sqrt{t(t+\tau_{1/2})} \right] + \omega_0 \tau_{1/2} \ln\left( \frac{\sqrt{t} + \sqrt{t+\tau_{1/2}}}{\sqrt{\tau_{1/2}}} \right), \qquad (21)$$

i. e. the phase $\psi$ diverges logarithmically with time.

For the ion momentum dependence on time given by Eq. (16) $k = 3/5$. In this case we have $\psi \to 2.804 \times \omega_0 \tau_{3/5}$. In order to fulfill the ion trapping conditions the laser pulse duration must be larger than $t_{las} > 2.804 \times \tau_{3/5}$.

The possibility of reaching an unlimited ion energy that is formally allowed by the mirror model adopted here is actually limited to a finite value when transparency effects are included (see Refs. [1, 14]). Two effects compete in determining the transparency of the accelerated and expanding foil: as the foil momentum increases, in its proper frame of reference the frequency of the laser pulse decreases in the ultrarelativistic regime proportionally to $1/2\gamma$ (i. e., from Eqs.(11, 16) as $t^{-3/5}$), while the foil surface density decreases (as $t^{-4/5}$). At fixed dimensionless pulse amplitude, the foil transparency depends on the ratio between the foil surface density and the pulse frequency

$$a_0 \leq \frac{\varepsilon_p}{\Lambda_y \Lambda_z} \left( \frac{1+\beta}{1-\beta} \right)^{1/2}, \qquad (22)$$

where $a_0 = eE_0 / m_e \omega_0 c$ and $\varepsilon_p = 2\pi n_0 l_0 e^2 / m_e \omega_0 c$ (see Ref. [24]). For the expanding foil this ratio tends to zero as $t^{-1/5}$, making the foil transparent to the pulse radiation. For the shell expanding in only one transverse dimension, the shell can be always opaque for the incident laser pulse.

If the condition (22) is not fulfilled, i.e. $a_0 > \varepsilon_p \left[(1+\beta)/(1-\beta)\right]^{1/2} / \Lambda_y \Lambda_z$, which can be rewritten as $E_0 > 2\pi e n_0 l_0 \left[(1+\beta)/(1-\beta)\right]^{1/2} / \Lambda_y \Lambda_z$, or $E_\perp > 2\pi e n l$, where $E_\perp$ is the laser field in the co-moving with the mirror frame of reference, the ion acceleration develops according to the first acceleration stage noticed in Ref. [1]. In this case all the electrons are displaced in the longitudinal direction, which results in the formation of the charge separation electric field $E_\| > 2\pi enl$ between the electron and ion layers. In this longitudinal electric field the ion momentum grows as $p_x = 2\pi e^2 nlt$. In a time of the order of $t_1 = \omega_{pe}^{-1}\left(a_0/2\varepsilon_p^2 \Lambda_y \Lambda_z\right)(m_\alpha/m_e)$ the ion energy reaches the value when the condition (22) becomes valid. Then the ion acceleration will follow the above considered RPDA scenario.

In order to find the optimal conditions for unlimited acceleration we consider this regime in more detail. We represent the laser pulse shape by the function $E(\psi)$ in terms of the phase $\psi$ given by Eq. (17). Searching out this function is the problem of optimization. Changing the independent variable from $t$ to $\psi$, we write Eq. (12) in the form

$$\frac{dp_x}{d\psi} = \frac{m_\alpha^2 \omega_0 c}{\pi_\perp^0} h^2(\psi) \frac{\Lambda^2}{(1+\beta)} , \qquad (23)$$

where we introduced a normalized intensity of the laser pulse

$$h^2(\psi) = \frac{v_E^2(\psi)\pi_\perp^0}{l_0 m_\alpha \omega_0^2 c} = \frac{E^2(\psi)\pi_\perp^0}{2\pi n_0 l_0 m_\alpha^2 \omega_0^2 c} . \qquad (24)$$

Here for simplicity we assume that the shell expansion is isotropic, i. e. $\pi_y^0 = \pi_z^0 = \pi_\perp^0$ and $\Lambda_y = \Lambda_z = \Lambda$, so that from Eqs (10) and (17) we have

$$\frac{d\Lambda}{d\psi} = \frac{\pi_\perp^0}{m_\alpha \omega_0}\left(\frac{1+\beta}{1-\beta}\right)^{1/2} . \qquad (25)$$

System of Eqs. (23, 25) can be cast in the form

$$\frac{d^2\Lambda}{d\psi^2} = h^2(\psi)\Lambda^2 , \qquad (26)$$

.

If we take the laser pulse amplitude to be equal to $E_0$ for $0 < \psi < \omega_0 t_{las} = \psi_m$ and to vanish for $\psi < 0, \psi > \psi_m$, then according to Eq. (24) the laser normalized amplitude, $h(\psi)$, is constant,

$$h_0 = \left( \frac{v_{E,0}^2 \pi_\perp^2}{l_0 m_\alpha \omega_0^2 c} \right)^{1/2}, \tag{27}$$

in the interval $0 < \psi < \psi_m$ and vanishes outside. The solution to Eq. (26) can be expressed in terms of the Weierstrass elliptic function, $\wp(u, \{g_2, g_3\})$,

$$\Lambda(\psi) = \frac{6}{h_0^2} \wp\left(\psi_* - \psi, \{0, g_3\}\right) \tag{28}$$

where $g_2 = 0$ and $g_3$ are constant determined by initial conditions

$$g_3 = \frac{h_0^4}{36} \left( \frac{2 h_0^2}{3} \Lambda_0^3 - \Lambda_0'^2 \right). \tag{29}$$

The Weierstrass elliptic function $\wp(u, \{g_2, g_3\}) = z$ gives the value of $z$ for which [20]

$$u = \int_\infty^z \frac{dt}{\sqrt{4t^3 - g_2 t - g_3}} dt. \tag{30}$$

According to Eq. (28), the function $\Lambda(\psi)$ becomes singular for the phase $\psi_*$ determined by the smallest positive solution to $\wp(\psi_*, \{0, g_3\}) = h_0^2 \Lambda_0 / 6$.

At $\psi \to \psi_*$ Eq. (28) yields

$$\Lambda(\psi) = \frac{6}{h_0^2 (\psi_* - \psi)^2} + \frac{3 g_3}{14 h_0^2} (\psi_* - \psi)^4 + O((\psi_* - \psi)^9). \tag{31}$$

The solution to Eq. (23) with the initial condition $p_x(\psi = 0) = 0$ gives a dependence of the momentum on the phase:

$$\left(m_\alpha^2 c^2 + p_x^2\right)^{1/2} + p_x = m_\alpha c + \frac{6 m_\alpha^2 \omega_0 c}{h_0^2 \pi_\perp^*} \left[\wp'(\psi_*, \{0, g_3\}) - \wp'(\psi_* - \psi, \{0, g_3\})\right]. \quad (32)$$

Here $\wp'(u, \{g_2, g_3\})$ is the derivative of the Weierstrass elliptic function with respect to the variable $u$. From this expression it follows that $p_x \propto (\psi_* - \psi)^{-3}$ at $\psi \to \psi_*$. From Eq. (17) we find the dependence of the phase on time, $(\psi_* - \psi) \propto t^{-1/5}$, in the limit $\psi \to \psi_*$. These relationships correspond to the time dependence of the momentum given by Eq. (16).

The ion acceleration can be effectively optimized by tailoring the laser pulse shape. Assuming the laser pulse shape to depend as

$$h(\psi) = h_0 (\psi_* - \psi)^m \quad (33)$$

we obtain the exact solution of Eq. (26)

$$\Lambda(\psi) = \frac{2(1+m)(3+2m)}{h_0^2 (\psi_* - \psi)^{2(1+m)}}. \quad (34)$$

Using this dependence and integrating Eq. (23) we obtain the dependence of the momentum on the phase

$$p_x = m_\alpha c \frac{\chi}{2} \frac{\left[\psi_*^{3+2m} - (\psi_* - \psi)^{3+2m}\right]\left[(2-\chi)(\psi_* - \psi)^{3+2m} + \chi \psi_*^{3+2m}\right]}{(1-\chi)(\psi_* - \psi)^{6+4m} + \chi\left[\psi_*(\psi_* - \psi)\right]^{3+2m}}, \quad (35)$$

where

$$\chi = 4(1+m)^2(3+2m)\left(\frac{m_\alpha\omega_0}{\pi_\perp^0 \psi_*^{3+2m}}\right). \tag{36}$$

For $m > -3/2$, the asymptotic for $\psi \to \psi_*$ is

$$p_x = \frac{m_\alpha\omega_0}{\pi_\perp^0}\frac{2(1+m)^2(3+2m)}{(\psi_* - \psi)^{3+2m}} + \ldots \tag{37}$$

Integration of Eq. (17) yields the time dependence of the phase $\psi(t)$. It reads

$$\frac{\chi^2\psi_*}{5+4m}\left[\left(\frac{\psi_*}{\psi_* - \psi}\right)^{5+4m} - 1\right] + \frac{(1-\chi)\chi\psi_*}{1+m}\left[\left(\frac{\psi_*}{\psi_* - \psi}\right)^{2(1+m)} - 1\right] + \left[1 + (1-\chi)^2\right]\psi = 2\omega_0 t, \tag{38}$$

or, asymptotically for $\psi \to \psi_*$ (provided that $m > -5/4$),

$$\psi = \psi_* - \left[\left(\frac{\pi_\perp^0}{m_\alpha\omega_0}\right)^2\frac{(5+4m)\omega_0 t}{8(1+m)^4(3+2m)^2}\right]^{-\frac{1}{5+4m}} + \ldots \tag{39}$$

This results in the momentum time dependence

$$p_x = m_\alpha c\left(\frac{t}{\tau_k}\right)^k + \ldots, \tag{40}$$

where the power $k = (3+2m)/(5+4m)$. We note that it satisfies the condition $k > 1/2$ for any $m$. The characteristic acceleration time $\tau_k$ is equal to

$$\tau_k = \frac{2}{(5+4m)\omega}\left[2(1+m)^2(3+2m)\frac{m_\alpha\omega_0}{\pi_\perp^0}\right]^{\frac{1}{3+2m}}. \tag{41}$$

The momentum per unit surface of the shell asymptotically depends on time as

$$p_x nl \propto t^{-(1+2m)/(5+4m)}. \tag{42}$$

If the power $-5/4 < m < -1/2$, the momentum per unit surface grows in time with the same asymptotic as the r.h.s. of the opaqueness condition Eq. (22). This means that the radiation pressure of the laser pulse with the shape being a power of the phase can accelerate an expanding shell in such a way that the shell remains opaque for the laser radiation.

## 4. Nonrelativistic limit

In the nonrelativistic limit, when $p_i = m_\alpha \dot{x}_i$, Eqs. (6, 7) are reduced to equations

$$\partial_{tt} x_i = \frac{\mathcal{P}}{m_\alpha \sigma_0} \varepsilon_{ijk} \partial_\eta x_j \partial_\zeta x_k, \tag{43}$$

which admit the exact solution

$$x = \frac{v_{E,0}^2 \tau_{ex}^2}{12 l_0} \left[ \left(1 + \frac{t}{\tau_{ex}}\right)^4 - 1 \right] \tag{44}$$

with

$$y = \Lambda(t)\eta, \quad z = \Lambda(t)\eta, \quad \Lambda(t) = 1 + \frac{t}{\tau_{ex}}, \tag{45}$$

where $\tau_{ex} = m_\alpha / \pi_\perp^0$ is an expansion time. Asymptotically for $t \to \infty$ the ion kinetic energy grows as

$$\mathcal{E}_\alpha \approx \frac{m_\alpha v_{E,0}^4}{18 l_0^2 \tau_{ex}^4} t^6 \tag{46}$$

and the ion areal density decreases as $nl \approx n_0 l_0 /(1 + t/\tau_{ex})^2$. Assuming the acceleration time is the laser pulse duration, $t_{las}$, and writing the laser pulse fluence as $w_{las} = cE^2 t_{las}/4\pi = I t_{las}$, where $I$ is the laser intensity, we find for the ion acceleration efficiency

$$\mathcal{K}_{eff} = \frac{cE^2 t_{las}^3}{18\pi m_\alpha c^2 n_0 l_0 \tau_{ex}^2} \equiv \frac{2 I t_{las}}{9 m_\alpha c^2 n_0 l_0} \left(\frac{t_{las}}{\tau_{ex}}\right)^2. \tag{47}$$

This efficiency is by a factor $(t_{las}/\tau_{ex})^2$ higher than in the case of non-expanding foil [1], which points towards a way for enhancing the efficiency of the fast ion generation required within the framework of the concept of Fast Ignition with laser accelerated ions [25].

As we see the efficiency enhancement requires the laser pulse duration to be larger than the foil expansion time. Assuming the expansion time to be of the order of the inverse growth rate of the Raleigh-Taylor instability, $(q_\perp v_E^2 / l_0)^{1/2}$ [9], with the wavelength of transverse perturbations equal to the inverse laser pulse waist, $q_\perp = 2\pi/w$, we find $t_{las}/\tau_{ex} = v_E t_{las}/(w l_0 / 2\pi)^{1/2}$. For $v_E = c a_0 (m_e/m_\alpha)^{1/2} (\omega_0/\omega_{pe})$, $a_0 = 100$, $l_0 = 0.1\lambda_0$ and $w \approx c t_{las}$ with $t_{las} \approx 100$ fs, it yields $v_E \approx 0.1 c$ and the factor $(t_{las}/\tau_{ex})^2 \approx 10 - 20$.

## 5. PIC simulations of the radiation pressure acceleration of a mass limited target

As an illustration of the realization of the RPDA regime in the interaction of a high intensity laser pulse with an expanding plasma shell we present the results of two-dimensional (2D) particle-in-cell (PIC) simulations of the dynamics of a mass limited target (MLT) in a strong laser field. We use the PIC code REMP [26]. As well known the MLT irradiated by strong laser light can behave similarly to the cluster. The Coulomb explosion of laser irradiated clusters is considered to be one of the basic mechanisms of ion acceleration [27]. When clusters of sufficiently small size are irradiated by a strong electromagnetic pulse, the ions can be accelerated together with the electrons by the RPDA mechanism up to an energy substantially higher than the energy achievable in the case of pure Coulomb explosion.

In Figs. 2 – 5 we present the results of the 2D PIC simulations of the interaction of an ultraintense laser pulse with a MLT. The simulation box has the size of $600\lambda \times 100\lambda$ with the mesh resolution of 20 cells per wavelength. The number of particles is equal to $10^4$. The target has the form of an ellipsoid in the $(x, y)$ plane with horizontal and vertical semi-axes equal to $1\lambda$ and $7.5\lambda$. It is initially localized at $x = 50\lambda$, $y = 0$. The target is made of hydrogen plasma with the proton-to-electron mass ratio equal to 1836. The electron density corresponds to the ratio $\omega_{pe}/\omega = 6$. A circularly polarized laser pulse is excited in the vacuum region at the left hand side of the computation region. The laser pulse has a super-Gaussian shape with a length of $l_x = 25\lambda$, a width of $l_y = 25\lambda$ and with the dimensionless amplitude equal to 125.

The value $a_0 = 125$ of the dimensionless amplitude of the laser pulse approaches the threshold above which it is necessary to take into account the influence of radiation friction on the dynamics of the electron component [28, 29]. It was noted in Refs. [1, 30] that, as the target is accelerated, the influence of radiation friction decreases. Intensity of radiation emitted by electron is given by

$$I = \frac{2e^2}{3m_e^2 c^3}\left(\frac{dp_\mu}{ds}\frac{dp_\mu}{ds}\right), \tag{48}$$

where $p_\mu$ is the electron 4-vector, $\mu = 0,1,2,3$, $s = \int dt/\gamma$ is the proper time.

In the circularly polarized electromagnetic wave, whose amplitude is equal to $E_0$ the electron energy losses in the rest frame are given by

$$\dot{\mathcal{E}}^{(-)} = -\frac{2e^4 E_0^2}{3m_e^2 c^3}\left[1+\left(\frac{eE_0}{m_e \omega_0 c}\right)^2\right], \tag{49}$$

and the linearly polarized wave they are

$$\dot{\mathcal{E}}^{(-)} = -\frac{e^4 E_0^2}{3m_e^2 c^3}\left[1+\frac{3}{8}\left(\frac{eE_0}{m_e \omega_0 c}\right)^2\right], \tag{50}$$

The electromagnetic wave can provide energy gain rate not higher than

$$\dot{\mathcal{E}}^{(+)} \approx eE_0 c = \omega_0 m_e c^2 a_0. \tag{51}$$

A condition of the energy balance, $\dot{\mathcal{E}}^{(+)} = -\dot{\mathcal{E}}^{(-)}$, yields the wave dimensionless amplitude for the threshold when radiation friction becomes important

$$a_{rad}^c \geq \left(\frac{3\lambda_M}{4\pi r_e}\right)^{1/3}. \tag{52}$$

in the case of the circularly polarized electromagnetic wave, and

$$a_{rad}^l \geq \left(\frac{4\lambda_M}{\pi r_e}\right)^{1/3}. \tag{53}$$

for linear polarization. For details see Refs. [29, 31]. Here $\lambda_M$ is the radiation wavelength in the rest frame of reference of the accelerated mirror, i.e. it is equal to $\lambda_M = \lambda_0\sqrt{(1+\beta)/(1-\beta)} \approx \lambda_0 2\gamma$, and $r_e = e^2/m_e c^2 \approx 2.8\times 10^{-13}$ cm is the classical electron radius. For the $\lambda_0 = 0.8\,\mu$m radiation wavelength, if $\gamma = 1$, i. e. $\lambda_M = \lambda_0$, we have $a_{rad}^c = 408$ and $a_{rad}^l = 713$, which corresponds to the laser intensity $4.5\times 10^{23}$ W/cm$^2$ and $7\times 10^{23}$ W/cm$^2$, respectively. For the relativistic factor $\gamma$ corresponding to an energy of 10 GeV, the threshold value of the dimensionless amplitude of the electromagnetic wave is above 1000, which is more than eight times larger than the amplitude used in the above numerical simulations.

As it is seen in the simulation, the laser pulse radiation pressure compresses the target in the longitudinal direction. The reflected light wavelength increases since the incident wave interacts with the receding relativistic mirror, as it is seen in Fig. 2, where a superposition of the ion density and the z-component of the electric field in the $(x, y)$ plane is shown for $t = 112.5 \times 2\pi / \omega$.

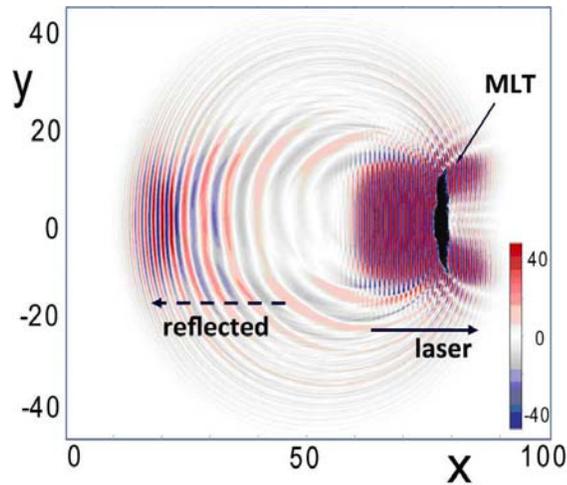

Figure 2. The laser pulse, reflected radiation and mass limited target shown as a superposition of the ion density distribution and the $z$-component of the electric field in the $(x, y)$ plane at time, $t = 112.5 \times 2\pi / \omega$. The length is normalized to the laser wavelength $\lambda$.

Figs. 3, 4 and 5 illustrate a time evolution of the laser pulse and target. In Fig. 3 (a) and (b) we show the $x$- and $z$- components of the electric field at $t = 100 \times 2\pi / \omega$. Figs. 3 (c) and (d) present the electron and ion density distribution in the $(x, y)$ plane. In Figs. 3 (a) and (b) we see a relatively long wavelength reflected radiation and the positive $x$-component of the electric field accelerating ions in the forward direction. The laser pulse expels a portion of the electrons in forward direction forming ultra short electron bunches in the region ahead of the target (see

also Refs. [32, 33]). The main part of the electrons and ions is compressed in the longitudinal direction. Behind the compressed shell we see a low-density plasma cloud.

The particles with the highest energy are localized in the high density shell as it is seen in Fig. 4, where the phase planes are shown for the electrons [Fig. 4 (a)] and for the ions [Fig. 4 (b)] at $t = 100 \times 2\pi / \omega$. The electron momentum is normalized to $m_e c$ and the ion momentum is normalized to $m_p c$.

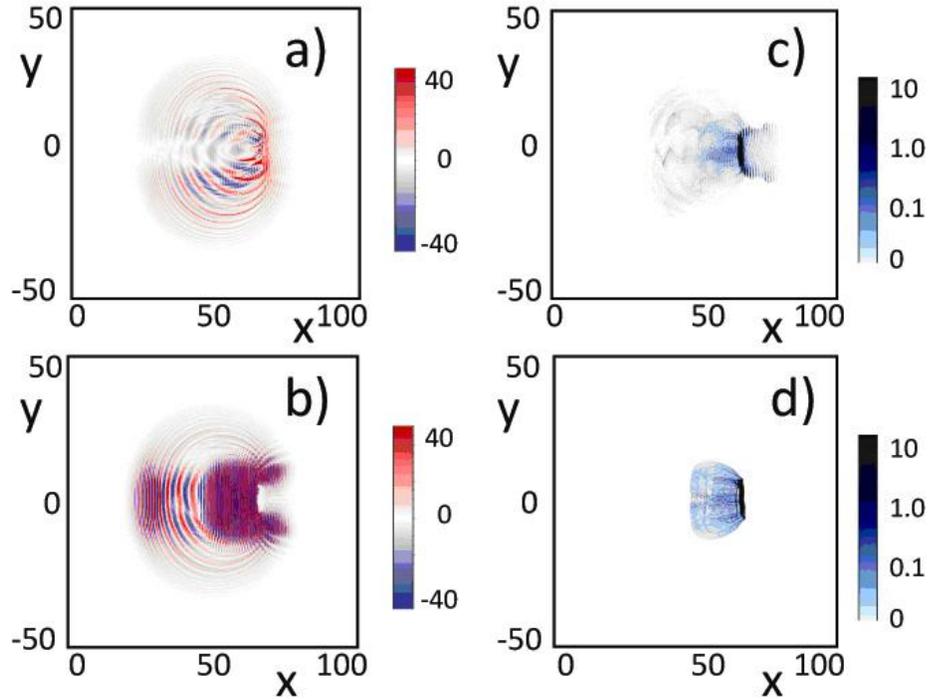

Figure 3. The electromagnetic field and mass limited target at $t = 100 \times 2\pi / \omega$ shown in the $(x, y)$ plane by a) the $x$-component of the electric field, b) the $z$-component of the electric field, c) the electron density, and d) the ion density.

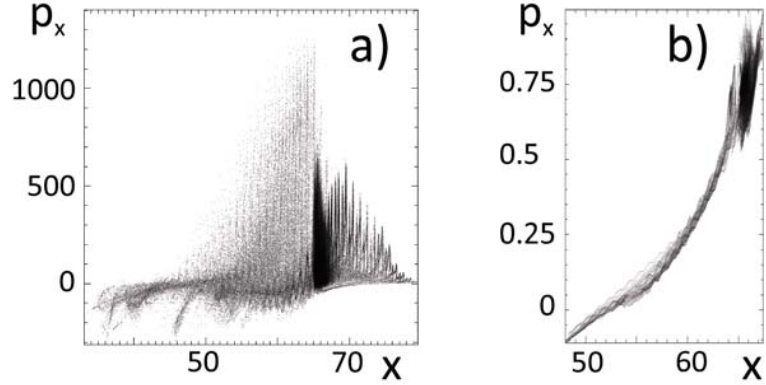

Figure 4. The phase planes of the electrons (a) and ions (b) at $t = 100 \times 2\pi/\omega$. The electron momentum is normalized to $m_e c$ and the ion momentum is normalized to $m_p c$.

At later time the target expands along the transverse direction as seen in Fig. 5, where the same quantities are shown as in Fig. 3 but for the time $t = 250 \times 2\pi/\omega$. This makes the laser pulse to interact with a thin dense shell expanding in the transverse direction in a regime close to one discussed above.

In Fig. 6 we present the electron and ion energy and the maximum ion density versus time. At the initial stage the ion density increases and then tends to zero. The electron and ion energies grow being of the same order of magnitude. From Fig. 6 we see that at time $t = 600 \times 2\pi/\omega$, when the accelerated shell approaches the right hand side boundary of the computation box, the protons reach the energy of 14 GeV and the electron energy is equal to 27 GeV. In the inset we show the ion energy spectrum at $t = 600 \times 2\pi/\omega$, which demonstrates the mono-energetic peak with the width of the order of 5%.

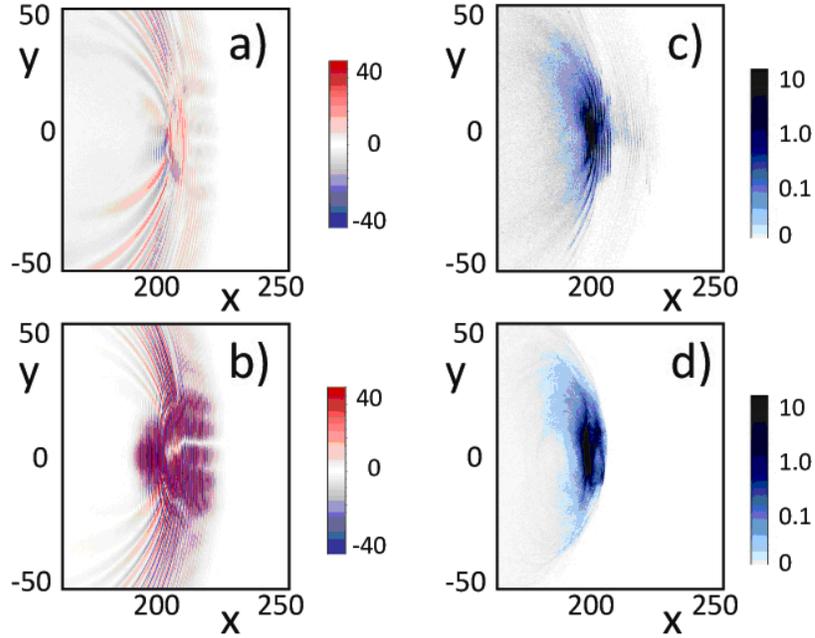

Figure 5. The electromagnetic field and mass limited target at $t = 250 \times 2\pi/\omega$ shown in the $(x,y)$ plane by a) the $x$-component of the electric field, b) the $z$-component of the electric field, c) the electron density, and d) the ion density.

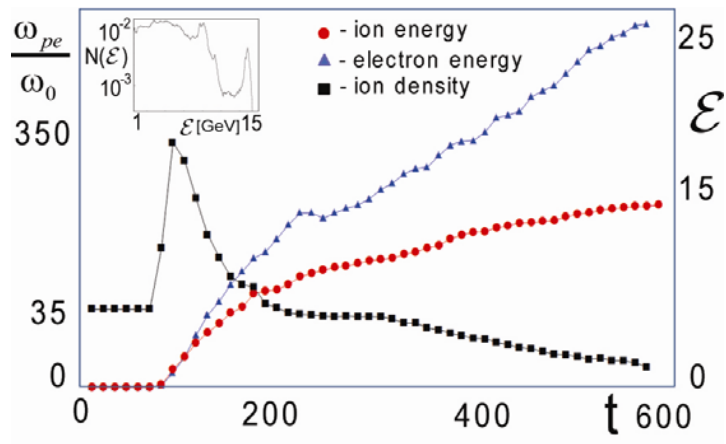

Figure 6. Electron and ion energy and the normalized Langmiur frequency corresponding to the ion density versus time. Inset: Ion energy spectrum at $t = 600 \times 2\pi/\omega$.

If we estimate the energy of the accelerated ions according to expression (11) for the simulation parameters assuming that no deformation of the target occurs, we find it to be of the order of 3 GeV. The enhancement of the fast ion energy is due to the target expansion in the transverse direction as discussed above.

## 6. PIC simulations of radiation pressure acceleration of ions in the laser pulse interaction with a thin foil target

In order to compare the ion acceleration by the laser pulse interacting with the mass limited target with the case when the laser interacts with a thin foil target we performed particle-in-cell simulations of the interaction of a circularly polarized electromagnetic pulse with a thin overdense plasma target consisting of electrons and protons (see also [29, 30]). The target has the form of an ellipsoid in the $(x, y)$ plane with horizontal and vertical semi-axes equal to $1\lambda$ and $50\lambda$. It is initially localized at $x = 50\lambda$, $y = 0$. The dimensionless amplitude of the laser pulse is $a = 125$, which corresponds to a laser intensity of $I = 3.12 \times 10^{22}$ W/cm$^2$ for a wavelength of $\lambda = 1\mu$m. The laser pulse has a super-Gaussian shape with a length of $l_x = 25\lambda$, a width of $l_y = 25\lambda$. The target is made of hydrogen plasma with the proton to electron mass ratio equal to 1836. The electron density corresponds to the ratio $\omega_{pe}/\omega = 6$.

Due to the interaction with the laser pulse, the target changes its shape and transforms into a cocoon, which traps the electromagnetic wave [Figs. 7 (a) – (d)], thereby providing ion acceleration over a distance longer than the Raleigh length. At the cocoon front a dense plasma clump is formed as was noticed in Ref. [9]. The clump propagates with a relativistic velocity. As a result, the coefficient of reflection of the electromagnetic wave from the supercritical plasma layer moving with a velocity approaching the speed of light progressively increases. It can be seen from Fig. 8 (b) that, at the instant of $t = 112.5 \times 2\pi/\omega$, the maximum proton energy is about 2.5 GeV, whereas the electron energy is about 2 GeV.

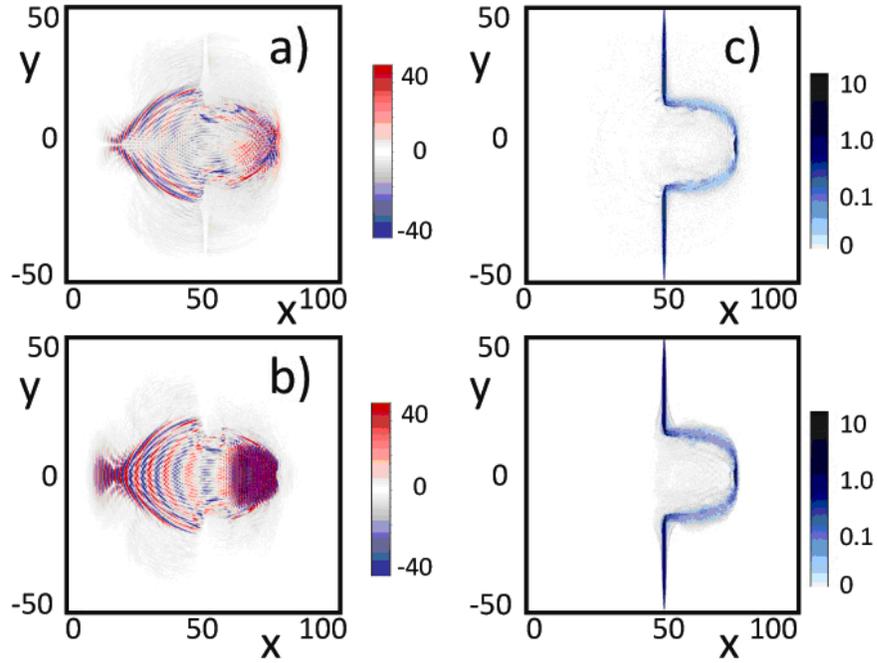

Figure 7. The electromagnetic field and thin foil target at $t = 112.5 \times 2\pi/\omega$ shown in the $(x, y)$ plane by a) the $x$-component of the electric field, b) the $z$-component of the electric field, c) the electron density, and d) the ion density.

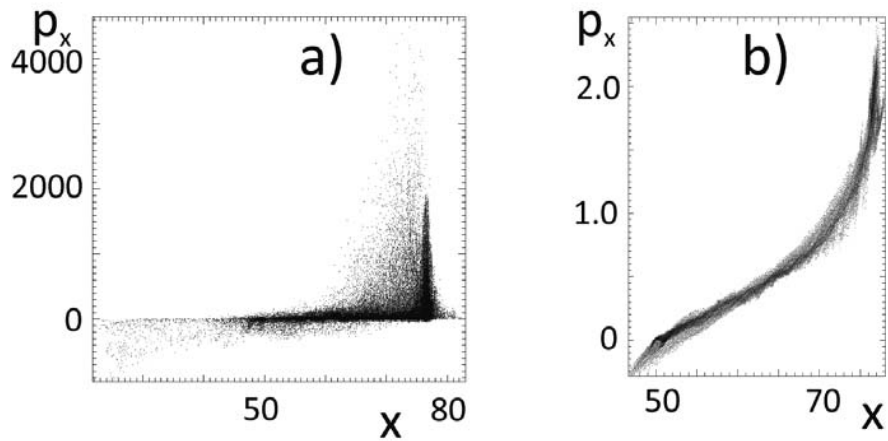

Figure 8. The phase planes of the electrons (a) and of the ions (b) at $t = 112.5 \times 2\pi/\omega$. The electron momentum is normalized to $m_e c$ and the ion momentum is normalized to $m_p c$.

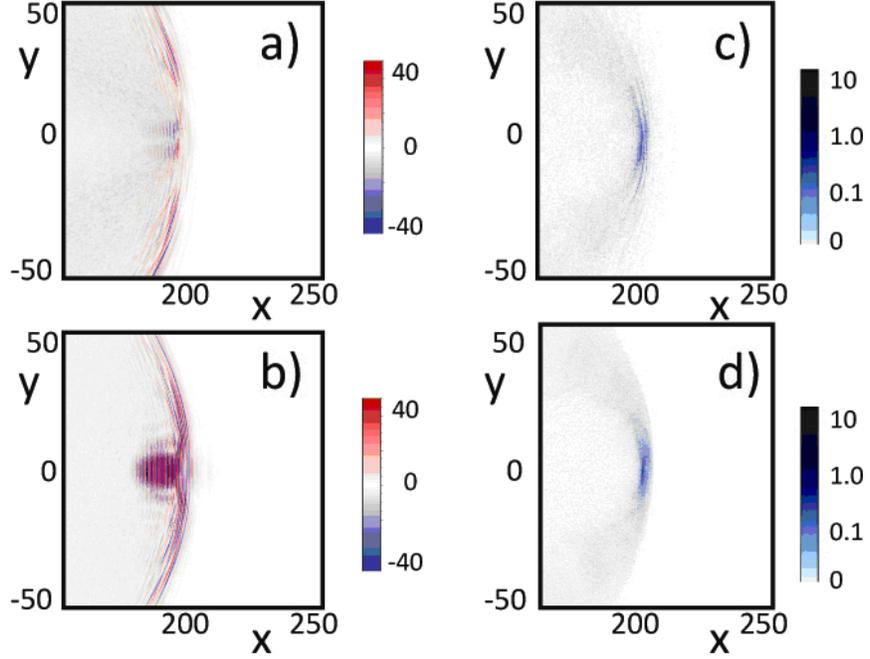

Figure 9. The electromagnetic field and thin foil target at $t = 250 \times 2\pi/\omega$ shown in the $(x, y)$ plane by a) the $x$-component of the electric field, b) the $z$-component of the electric field, c) the electron density, and d) the ion density.

At later time the accelerating part of the foil target evolves to a transversely expanding shell, similarly to a mass-limited target (see Fig. 9, where the same quantities as in Fig. 7 are shown but for the case of the foil and for $t = 250 \times 2\pi/\omega$). The laser pulse transforms the target into a new one thus changing the regime of its interaction, closer to the process described by our model and demonstrated above in the simulations with MLT. At the instant $t = 250 \times 2\pi/\omega$, the maximum proton energy is about 8.5 GeV.

## 7. Conclusion

In conclusion, the transverse expansion of a thin shell accelerated in the RPDA regime results in the increase of the ion energy and the acceleration efficiency at the expense of decreasing number of particles. In the relativistic limit, the ions become phase-locked with respect to the electromagnetic wave, which is the indication of an unlimited acceleration. This effect and the use of optimal laser pulse shape provide a new approach for great enhancing the energy of laser accelerated ions. Carried out [16] and forthcoming laboratory experiments on the laser plasma interaction in the RPDA regime will contribute to the development of the laboratory astrophysics discipline [34, 35] and to studying of the "Light Sail" mechanism for spacecraft propulsion.

## Acknowledgments

We acknowledge a partial support of this work from the MEXT of Japan, Grant-in-Aid for Scientific Research, projects 20244065 & 19740252 and ELI-PP contract number 212 105.